# The Ethics of Generative AI in Anonymous Spaces: A Case Study of 4chan's /pol/ Board

Parth Gaba[1] and Emiliano De Cristofaro[2]
[1]Valley Christian High School   [2]University of California, Riverside

## Abstract

This paper presents a characterization of AI-generated images shared on 4chan, examining how this anonymous online community is (mis-)using generative image technologies. Through a methodical data collection process, we gathered 900 images from 4chan's /pol/ (Politically Incorrect) board, which included the label "/mwg/" (memetic warfare general), between April and July 2024, identifying 66 unique AI-generated images. The analysis reveals concerning patterns in the use of this technology, with 69.7% of images including recognizable figures, 28.8% of images containing racist elements, 28.8% featuring anti-Semitic content, and 9.1% incorporating Nazi-related imagery.

Overall, we document how users are weaponizing generative AI to create extremist content, political commentary, and memes that often bypass conventional content moderation systems. This research highlights significant implications for platform governance, AI safety mechanisms, and broader societal impacts as generative AI technologies become increasingly accessible. The findings underscore the urgent need for enhanced safeguards in generative AI systems and more effective regulatory frameworks to mitigate potential harms while preserving innovation.

***Disclaimer:*** *This repor contains unsafe language and imagery that might be highly offensive to some readers, such as anti-semitic content. We include them to highlight the risks of AI-generated content and raise awareness about their potential misuse. Reader discretion is advised.*

## Introduction

The rapid advancement and democratization of generative AI technologies have transformed the landscape of digital content creation [5]. Tools like DALL·E [2], StyleGAN [14], and other image generation models enable individuals with minimal technical expertise to create sophisticated visual content that was previously possible only for skilled graphic designers and artists. While this technological revolution has spawned numerous positive applications across art, design, and entertainment, it has simultaneously created new vectors for potential misuse.

4chan [1], an anonymous imageboard platform founded in 2003, has evolved [4] from its origins as an anime and manga discussion forum into a complex ecosystem of diverse communities. Known for both its creative contributions to Internet culture and its controversial content, 4chan—particularly its/pol/ (Politically Incorrect) board—has developed a reputation [4,13] as an incubator for extreme ideologies, offensive content, and harmful memes. The platform's emphasis on anonymity, coupled with nearly non-existent content moderation, creates an environment where controversial and potentially harmful content can proliferate without accountability and "spread" on mainstream platforms.

This paper investigates the intersection of generative AI technologies and the 4chan community, focusing on how AI-generated images are created, shared, and repurposed across the platform. Through systematic data collection from the /pol/ board and the "/mwg/" (memetic warfare general) board, we aim to characterize the nature, themes, and potential impacts of AI-generated content in this online ecosystem. The research addresses critical questions about how emerging technologies are being deployed in anonymous online spaces, the challenges these developments pose for content moderation and platform governance, and the broader implications for technology ethics and policy.

By understanding how communities like 4chan utilize generative AI for image creation, we can better anticipate challenges, develop appropriate safeguards, and inform policy discussions surrounding the responsible development and deployment of these powerful technologies. This research contributes to the growing body of literature on online extremism, technological misuse, and the social implications of artificial intelligence in digital communities.

## Background

### What is 4chan?

Christopher Poole launched 4chan as an anonymous imageboard in 2003. Initially created for discussions on anime and manga, it has evolved into a platform hosting a wide range of topics, including technology, politics, and random discussions. The site is divided into various boards, each dedicated to specific subjects, such as technology (/g/), random (/b/), and politically incorrect (/pol/). The platform's emphasis on anonymity has fostered a unique culture characterized by both creativity and controversy.

4chan operates with a design centered around ephemerality and anonymity, where users typically post without registration and are identified only by the time and content of their post, rather than a persistent username or profile. This lack of identity persistence fosters a culture of unaccountability and rapid content generation. Research has linked this structure—particularly in boards like /pol/

(Politically Incorrect)—to the proliferation of offensive, conspiratorial, and extremist material, including hate speech and racist memes [4].

4chan has been a significant incubator for internet memes, influencing various online subcultures. Memes like "Pepe the Frog" [17] and terms such as "Kek" [4] have origins that can be traced back to 4chan's boards. However, the platform has also been associated with the spread of hate speech and extremist ideologies, particularly on the /pol/ board, which has been identified as a space where far-right content is prevalent.

**Related Work**

Academic studies have examined 4chan's role in online culture, focusing on its anonymity, ephemerality, and impact on meme propagation. Bernstein et al. (2011) [1] conducted an analysis of 4chan's /b/ (random) board, highlighting its unique structure and the dynamics of user interactions. The study utilized a dataset comprising millions of posts to analyze user behavior and content trends.

Another study by Nissenbaum and Shifman [10] explored internet memes as contested cultural capital, using /b/ as a case study. They examined how memes function within the community and their broader cultural implications. Also, Hine et al. [4] focused on 4chan's /pol/ (Politically Incorrect) board, analyzing over 8 million posts to examine how anonymity and ephemerality contribute to the proliferation of hate speech, conspiracy theories, and politically extreme content. Their findings show that content from /pol/ frequently spreads to other platforms, such as Twitter and Reddit.

Zannettou et al. [15] further investigated 4chan's influence by tracking the origins and diffusion of memes across fringe and mainstream platforms. Their study identified /pol/ as a key source of memes—many with racist or conspiratorial themes—that later gain traction beyond 4chan's ecosystem.

Recent research has documented the emerging threat of AI-enabled harassment, with Marchal et al. [9] analyzing approximately 200 incidents of generative AI misuse between January 2023 and March 2024. Their findings reveal that the manipulation of human likeness and the falsification of evidence are prevalent tactics that require minimal technical expertise, significantly expanding the toolkit available to malicious actors on anonymous platforms like 4chan.

**What is Generative AI for Images?**

Generative AI for images allows the creation of new images based on patterns learned from large datasets. These models, such as Generative Adversarial Networks (GANs) and Variational

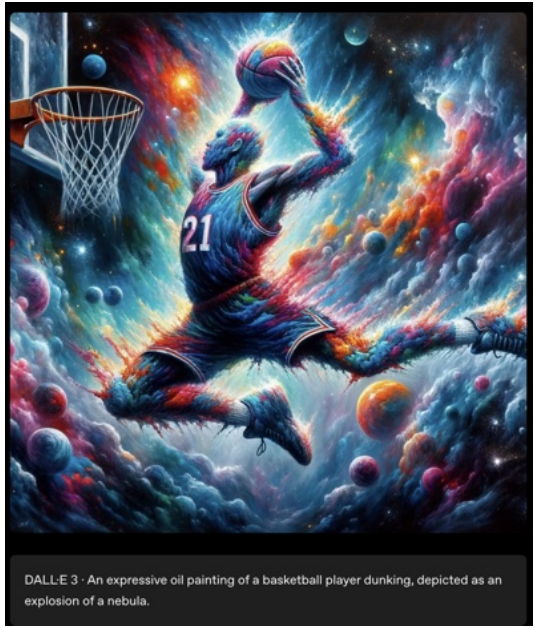

Figure 1: An image generated by DALL·E using a prompt describing an oil painting of a basketball player dunking amidst a nebula.

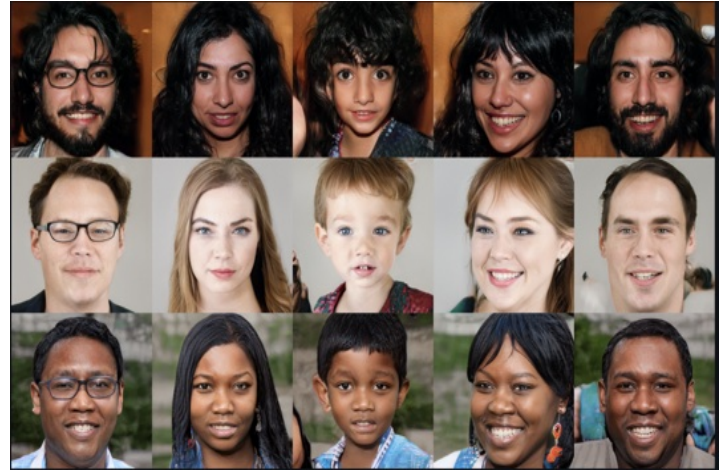

Figure 2: Image generated by StyleGAN to showcase highly realistic and detailed human faces.

Autoencoders (VAEs), can generate realistic or stylized images by learning from extensive datasets. GANs, introduced by Goodfellow et al. (2014), consist of two neural networks—the generator and the discriminator—that work in tandem to produce images indistinguishable from real ones.

Recent advancements include models like DALL·E, developed by OpenAI, which generates images from textual descriptions, showcasing the capability of AI to create novel visuals based on language inputs. There has also been further progress with new text-to-image models being integrated into foundation models like ChatGPT, Gemini, and Llama. However, we focus on DALL·E here because it has been more commonly discussed and used on /mgw/. These models have applications in art, design, and entertainment, enabling the creation of images that range from photorealistic to highly abstract.

**Systems Used in Generative AI for Images**

Several notable systems have been developed for image generation:

- **DALL·E**: Developed by OpenAI, DALL·E generates images from textual descriptions, allowing for the creation of imaginative visuals based on user-provided prompts. As illustrated in Figure 1, DALL·E can create surreal and expressive visuals, such as an oil painting of a basketball player dunking amidst a nebula, showcasing the system's ability to create imaginative and surreal visuals [2].

- **StyleGAN**: Developed by NVIDIA, StyleGAN is renowned for producing high-quality, photorealistic images, particularly of human faces. It offers fine-grained control over various image attributes, enabling the generation of diverse and detailed visuals. Figure 2 demonstrates this by displaying lifelike synthetic portraits of individuals [14].
- **memeBot**: Proposed by Sadasivam et al. (2020), memeBot is an approach to automatic image meme generation. It treats meme creation as a translation process, using an encoder-decoder architecture to generate image memes from textual input [12].

## How 4chan Users Discuss and Use Generative AI

Within the /pol/ board in 4chan, the /mwg/ (memetic warfare general) tag serves as a hub for discussions about AI-generated images, focusing on tools, techniques, and trends in the field. Users frequently share AI-generated artwork, discuss model performance, and provide guidance on how to fine-tune models for specific outputs. Through manual observation, we found that topics often include:

- Comparing different generative AI models (e.g., memeBot, StyleGAN, and DALL·E).
- Sharing guides on fine-tuning AI models for specific aesthetics.
- Debating AI-generated content, including deepfakes and NSFW material.
- Spreading Propaganda

Figure 3 provides an example of such a post from the /mwg/ tag, highlighting how these discussions and techniques are put into practice through shared AI-generated content. More precisely, the post discusses how to create AI-generated content, including propagandistic imagery featuring extremist symbols, Nazi-era uniforms, and fascist iconography deliberately integrated with contemporary meme culture to normalize and spread extremist ideologies through seemingly innocuous formats.

Overall, 4chan users interact with generative AI in a variety of ways, which we review next.

## Generative AI in Memes: Entertainment, Offensiveness, and Detection Challenges

Generative AI has been increasingly utilized for creating memes and humorous content. The accessibility of these tools has enabled users to experiment with AI-generated images for entertainment purposes. However, this democratization of image generation has also led to the creation of offensive or harmful content. The "Hateful Memes Challenge" introduced by Facebook AI Research specifically aimed to develop better detection systems for multimodal hate speech by creating a dataset of over 10,000 memes that combine benign images with hateful text or vice versa. The challenge revealed critical limitations in existing AI detection methods, as models struggled to understand the contextual relationship between visual and textual elements that make memes hateful. Key challenges include detecting implicit bias, understanding cultural context and internet slang, identifying visual dog whistles and coded language, and recognizing when seemingly innocent images become harmful when paired with specific text. This highlights the challenges in detecting hate speech in multimodal memes, underscoring the potential for misuse of generative AI in spreading offensive material [6,7].

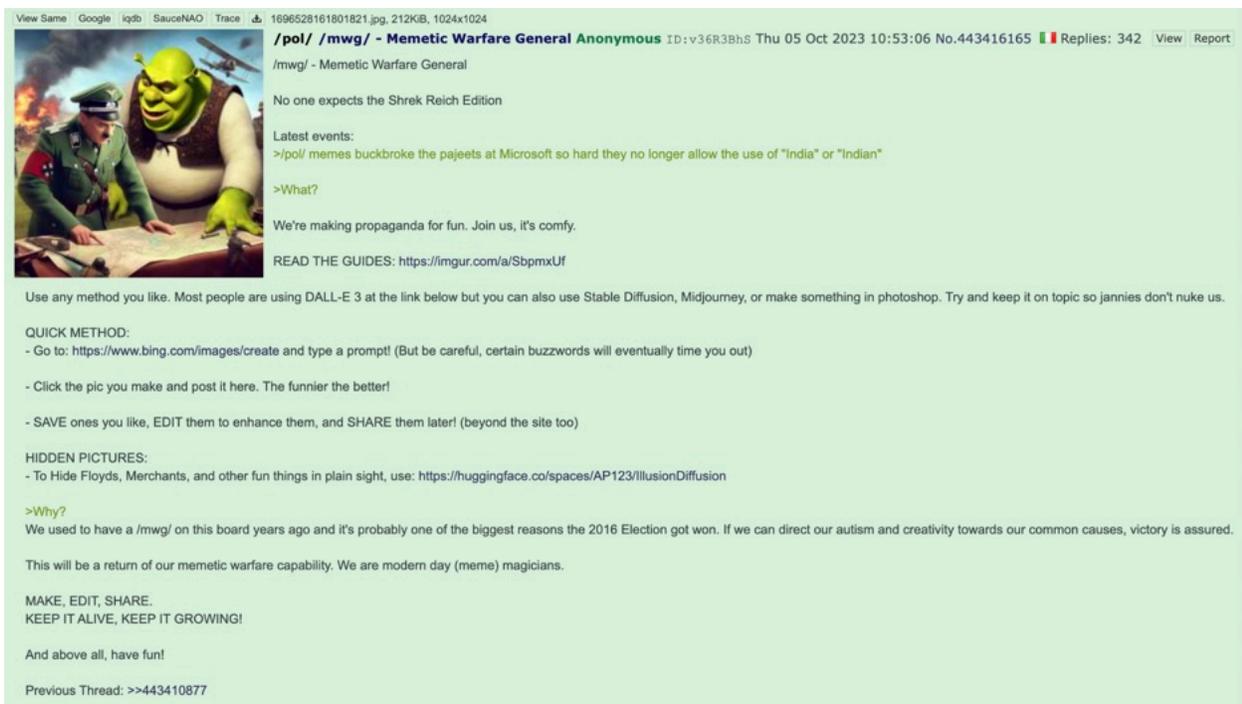

**Figure 3**: A post from 4chan's /pol/ board with the /mwg/ (memetic warfare general) tag, discussing not only AI tools and techniques but also how to actively create and share AI-generated content, including propagandistic imagery featuring historical extremist symbols.

## Data Collection Methodology

### Initial Attempt: Scraping 4chan_AI_Terror Twitter Account

The first attempt at data collection involved using Selenium to scrape images from the Twitter account @4chan_AI_Terror. This account is known for having content from 4chan's /pol/ board, including memes and AI-generated imagery. The goal was to extract images and text from the "Posts" tab, but this approach was unsuccessful due to Twitter's dynamic content loading and anti-bot measures. Instead, the focus shifted to the "Media" tab, which displayed only images, making extraction more straightforward.

We collected over 800 images from April to July 2024. The process involved:

- Configuring a Selenium WebDriver to log into Twitter.
- Navigating to the account's Media tab.
- Scrolling through the page and extracting image URLs.
- Downloading images

A relevant section of the extraction code:

```python
import os
import time
import requests
from selenium import webdriver
from selenium.webdriver.common.by import By
from bs4 import BeautifulSoup

def configure_driver():
    options = webdriver.ChromeOptions()
    options.add_argument("--start-maximized")
    driver = webdriver.Chrome(options=options)
    return driver

def download_image(img_url, save_path):
    img_data = requests.get(img_url).content
    with open(save_path, 'wb') as handler:
        handler.write(img_data)

driver = configure_driver()
driver.get('https://twitter.com/4chan_AI_Terror/media')
time.sleep(5)

soup = BeautifulSoup(driver.page_source, 'html.parser')
images = soup.find_all('img', {'src': True})

for i, img in enumerate(images):
    img_url = img['src']
    save_path = os.path.join('twitter_data/images', f'image_{i}.jpg')
    download_image(img_url, save_path)

driver.quit()
```

This approach ensured a consistent retrieval of images while minimizing redundant downloads.

**Attempting Data Collection from 4plebs Archive**

After collecting images from Twitter, we attempted to scrape data from **4plebs**, an archive site that stores historical posts from 4chan. The /pol/ board's archived threads were targeted for additional images. However, this approach failed due to *4plebs' strict anti-scraping measures*, which blocked Selenium-based automation.

**Direct Data Collection from 4chan's /mwg/ Tag Within /pol/ Board**

Since 4plebs did not allow automated data collection, the next step involved directly retrieving data from **4chan's threads tagged as /mwg/** (memetic warfare general), which is dedicated to AI-generated content. Unlike 4plebs, 4chan provides JSON-formatted data for threads, making extraction easier. We received JSON-formatted data from researchers Emiliano De Cristofaro and Jeremy Blackburn.

Over **100 images** were collected from /mwg/ within the same time period (April–July 2024) to ensure consistency. The process involved:

- Accessing JSON data
- Parsing thread responses to identify image links.
- Downloading images

**Image Similarity Analysis and AI-Generated Image Detection**

To compare images between **/mwg/** tag and 4chan_AI_Terror Twitter Account, perceptual hashing (pHash) was applied. This technique helped identify whether AI-generated images were shared across both platforms. The analysis revealed similarities, suggesting that AI-generated content was being disseminated in various contexts.

A relevant code excerpt for perceptual hashing:

```python
from PIL import Image
import imagehash
import os

def get_all_files(directory):
    return [os.path.join(directory, f) for f in os.listdir(directory) if not f.startswith('.')]

def find_similar_images(source_dir, compare_dir, threshold=5):
    similar_images = []
    for source_img_path in get_all_files(source_dir):
```

```
    hash_source = imagehash.phash(Image.open(source_img_path))
    for compare_img_path in get_all_files(compare_dir):
        hash_compare = imagehash.phash(Image.open(compare_img_path))
        if abs(hash_source - hash_compare) < threshold:
            similar_images.append(compare_img_path)
return similar_images
```

Following the similarity analysis, **all 900 collected images were processed to identify AI-generated content**. Initially, the **PIL** library was used for detection; however, it proved ineffective, as it falsely classified screenshots and other non-AI images. We implemented an API-based approach for higher accuracy.

The Hugging Face **AI-image-detector** model was used, which assigned probability scores to images that were artificially generated. The model classified images with a confidence score above 0.85 as AI-generated. This filtering process reduced the dataset to **100 AI-generated images**, which were then analyzed in further detail.

A snippet of the AI image detection code:

```
import requests
import os
import shutil

API_URL = "https://api-inference.huggingface.co/models/umm-maybe/AI-image-detector"
headers = {"Authorization": "Bearer YOUR_HUGGINGFACE_TOKEN"}

def query(filename):
    with open(filename, "rb") as f:
        data = f.read()
    response = requests.post(API_URL, headers=headers, data=data)
    return response.json()

source_folder = 'twitter_data/images/'
destination_folder = 'twitter_data/AI-Generated-Images/'

authentic_images = []
for image in os.listdir(source_folder):
    image_path = os.path.join(source_folder, image)
    output = query(image_path)
```

```
for item in output:
    if item.get("label") == 'artificial' and float(item.get("score", 0)) > 0.85:
        shutil.copy(image_path, destination_folder)
```

This final step ensured that only **high-confidence AI-generated images** were included in the final dataset for analysis. The combined methodology provided a **diverse and well-validated** dataset of AI-generated images sourced from 4chan and Twitter.

The combined methodology provided a diverse and well-validated dataset of AI-generated images sourced from 4chan and Twitter. All relevant code used in this project, including scraping scripts, perceptual hashing functions, and image classification tools, is available on GitHub: https://github.com/parthnonigaba/The-Ethics-of-Generative-AI-in-Anonymous-Spaces-A-Case-Study-of-4chan-s-pol-Board-Code.

## Characterization of AI-Generated Content on 4chan

### Overview of Themes and Content Categories

Upon analyzing the 66 AI-generated images collected from 4chan and Twitter, several recurring themes and content categories emerged. Each image was manually reviewed, with a short description written to summarize its content and tone. Based on these descriptions, we grouped images into thematic categories through qualitative coding, allowing for overlap when applicable (https://docs.google.com/spreadsheets/d/16SWYXYcQeFDtZDlm77f9pAma7UNxfNC5uP3D7YziDLs/edit?usp=sharing). This process helped identify patterns in the types of content being generated and shared.

**Extremist and Offensive Imagery**

- A significant portion (30%) of the dataset featured offensive or ideologically charged content, including anti-Semitic imagery with caricatures and Jewish stereotypes. Figure 4 exemplifies this, mocking Jewish families in a Pixar-style parody.
- References to Hitler and Nazi iconography were notable (10%), often shown in AI-generated images blending historical figures with meme culture. Figure 5 illustrates this by turning Hitler into a retro meme.

**Political Commentary and Satire**

- Many (70%) images convey strong far-right political views, often parodying social figures or criticizing progressive movements.

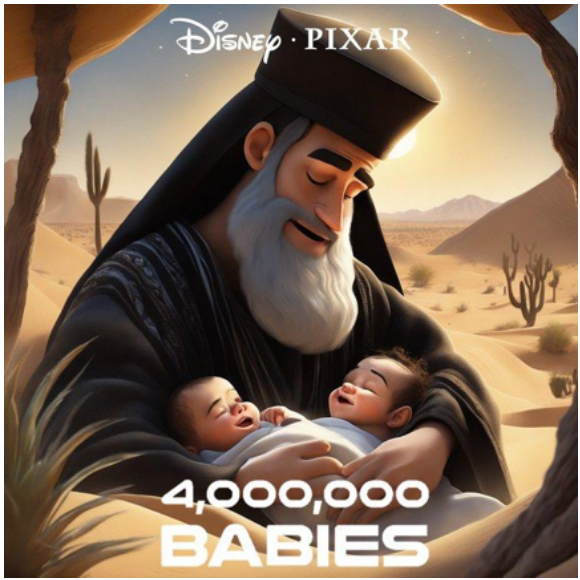

**Figure 4:** An image mimicing a Disney-Pixar movie poster, showing an elderly man in Jewish religious attire holding two babies in a desert. "4,000,000 Babies" is likely intended to mock Jewish families with an antisemitic stereotype.

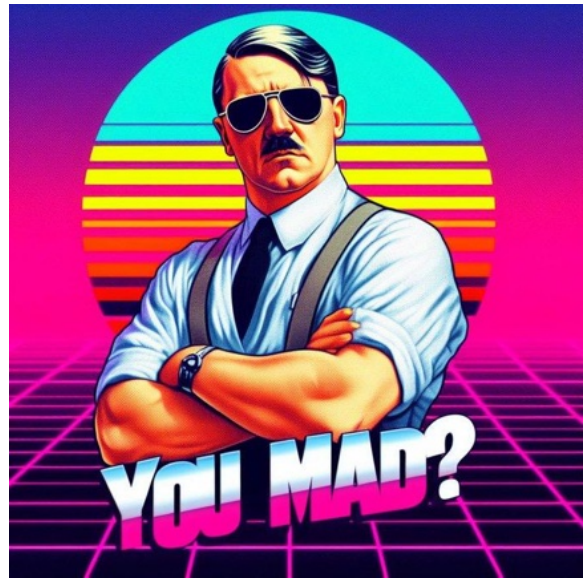

**Figure 5:** An image with Adolph Hitler wearing sunglasses and suspenders, styled like a 1980s poster with the caption "You Mad?" underneath.

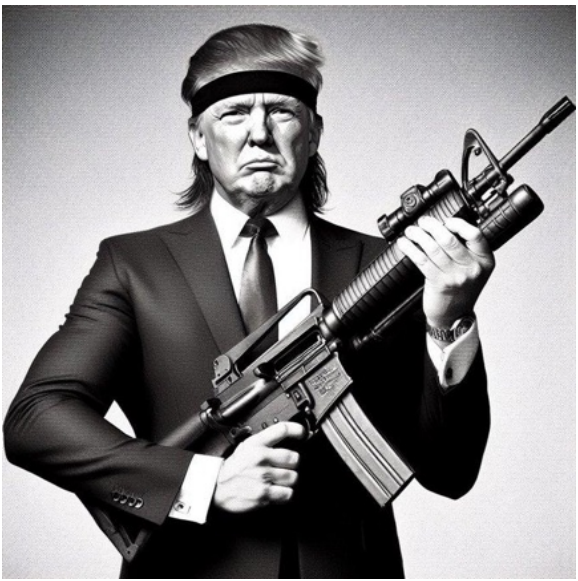

**Figure 6:** An image showing Trump holding an AK, exemplifying the use of AI-generated images to put public figures in controversial contexts.

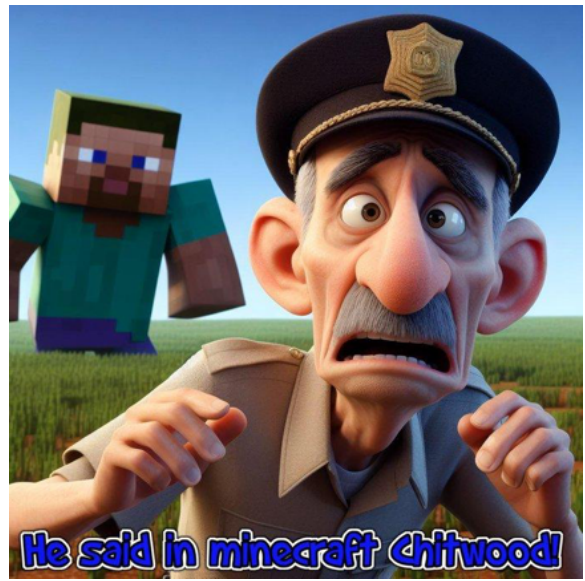

**Figure 7:** An image representing a response to a man who threatened Sheriff Mike Chitwood, ending the statement with "in Minecraft." Chitwood is depicted as a scared elderly white police officer with Steve from Minecraft in the background, implying that, since it was stated "in Minecraft," the man should not be arrested.

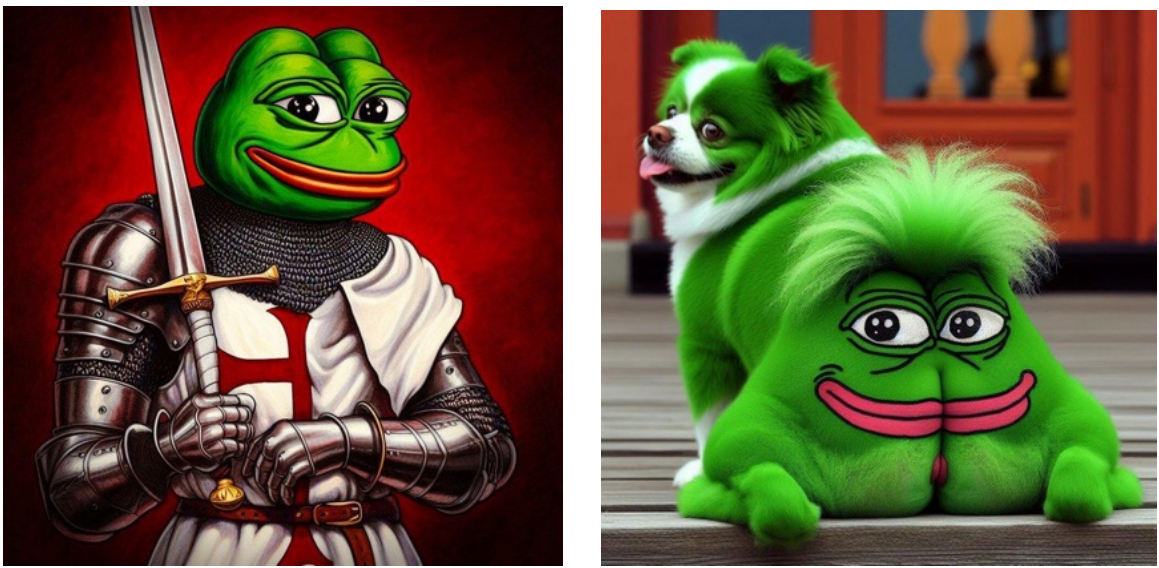

**Figure 8:** Examples of Pepe the Frog meme variants, specifically, as a knight and as a pet, illustrating how AI generates new versions of popular memes in unexpected ways.

- Some (10%) AI-generated images show politicians in exaggerated or demeaning ways, reflecting the political satire culture of the /pol/ board. Figure 6 exemplifies this by placing Trump in a militarized pose, heightening the symbolism.

**Meme Culture and Humor:**

- A considerable portion (20%) of the images are part of broader internet meme culture. These memes often utilize AI-generated art to enhance absurdist or surreal humor. Figure 7 reflects this trend through its parody of law enforcement and Minecraft culture.
- There are also playful or non-political AI-generated memes, though these are less frequent (5%) compared to more charged content. Figure 8 shows non-political meme variants like Pepe as a knight or pet, highlighting creative reinterpretations

**Common Visual and Textual Elements**

Several consistent visual and textual patterns were observed across the dataset, as detailed below.

Symbolism and Imagery:

- Frequent use of symbols like swastikas, Pepe the Frog variants, and "Happy Merchant" caricatures.
- Distorted or exaggerated human faces that align with the capabilities and limitations of StyleGAN-generated images.

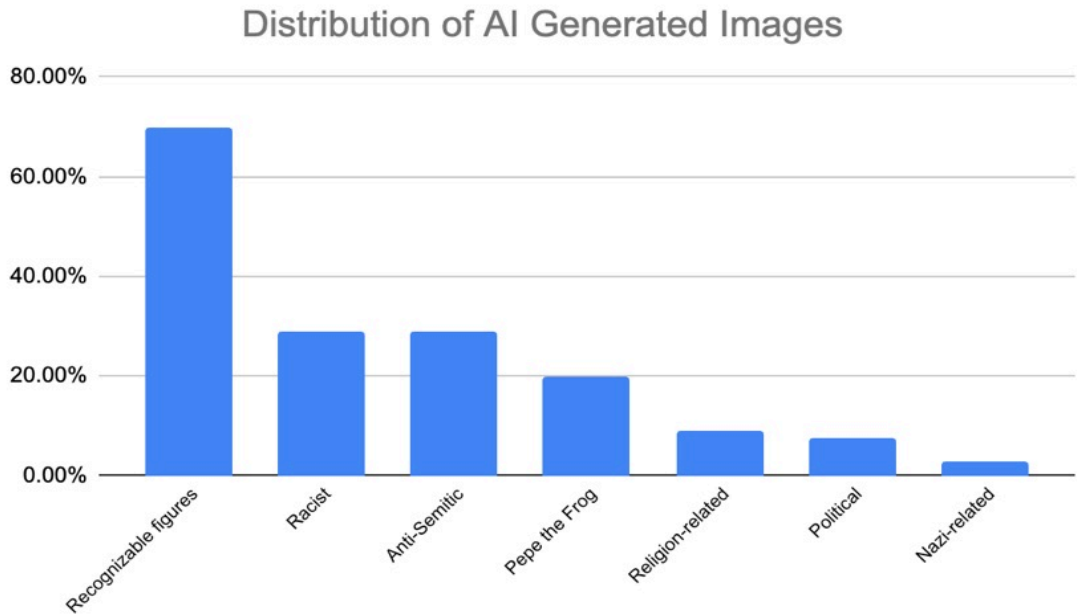

.

**Figure 9**: Distribution of AI-generated images in our dataset, showing the substantial presence of extremist content while also reflecting the platform's broader culture of meme-making and artistic experimentation.

Language and Text Overlays:

- Overlaid text often includes slurs, inflammatory language, or coded references to extremist ideologies.

Phrases like "based" and "red-pilled" signaling alignment with certain online subcultures.

### Quantitative Breakdown of Themes

A manual review of the 66 AI-generated images yielded the thematic distribution reported in Figure 9. (categories are non-exclusive, meaning images can belong to multiple categories)

## Conclusion

The comprehensive analysis of AI-generated images on 4chan reveals concerning patterns in how generative AI technologies are being utilized within anonymous online communities. The findings demonstrate that these powerful tools are frequently employed to create offensive, extremist, and potentially harmful content, with significant proportions of the analyzed images containing racist (28.8%), anti-Semitic (28.8%), and Nazi-related (9.1%) imagery. The prevalence of recognizable figures (69.7%) and culturally significant symbols like Pepe the Frog (19.7%) indicates strategic manipulation of popular cultural references to spread ideological messages.

**Discussion and Opinion**

The democratization of image generation technology presents a double-edged sword for society. While expanded creative capabilities offer tremendous potential for artistic expression and innovation, the findings demonstrate how easily these tools can be weaponized for harmful purposes. The ability to generate photo-realistic or stylized images of public figures in compromising or offensive contexts raises serious ethical concerns about misinformation, harassment, and the erosion of public trust.

Current content safeguards in generative AI models appear insufficient to prevent the creation of harmful imagery. Despite most commercial platforms implementing filters for explicit content and hateful imagery, the research shows that users can readily circumvent these protections through various prompt engineering techniques or by utilizing less restricted models. This suggests that technical solutions alone may be inadequate to address the full spectrum of potential misuse.

I believe additional safeguards are urgently needed, including more robust detection of coded language and symbols associated with extremist ideologies, improved identification of attempts to generate offensive content, and potentially, post-generation screening of images for harmful elements that may have bypassed initial filters.

From a societal perspective, the proliferation of AI-generated offensive imagery presents troubling implications for online discourse, marginalized communities, and information integrity. These technologies can amplify existing challenges in content moderation, accelerate the spread of extremist viewpoints, and potentially normalize harmful stereotypes and ideologies.

The development of comprehensive policy frameworks and potential legislation regarding generative models merits serious consideration. While avoiding overregulation that might stifle innovation, reasonable guardrails could include mandatory safety evaluations before public release of powerful generative models, transparency requirements regarding AI-generated content, and clearer accountability mechanisms for platforms that host such content.

**Future Research Directions**

Moving forward, several promising research avenues emerge from the work:

1. Scaling the analytical approach to examine larger datasets across multiple platforms, providing comparative insights into how different online communities utilize generative AI

2. Developing more sophisticated automated detection methods for identifying AI-generated extremist content, potentially employing machine learning techniques to recognize patterns indicative of harmful imagery
3. Conducting longitudinal studies to track the evolution of AI-generated content on platforms like 4chan as generative technologies advance
4. Investigating the effectiveness of various intervention strategies, from technical safeguards to policy approaches and educational initiatives
5. Examining the psychological and social impact of exposure to AI-generated extremist content on both platform users and broader society

In conclusion, the research underscores the urgent need for multidisciplinary collaboration between technology developers, platform operators, policymakers, and researchers to address the challenges posed by the misuse of generative AI. By understanding how these technologies are being utilized in spaces like 4chan, we can work toward more effective oversight mechanisms that balance innovation with responsible deployment, ultimately ensuring that these powerful creative tools benefit society while minimizing potential harm.

**Acknowledgments.** We wish to thank Prof. Jeremy Blackburn for providing help and assistance with 4chan imagery.

8. Kotousov, Gleb A., and Sergei L. Lukyanov. "ODE/IQFT Correspondence for the Generalized Affine $\mathfrak sl(2)$ Gaudin Model." *arXiv [Hep-Th]*, 2021, http://arxiv.org/abs/2106.01238.

9. Marchal, Nahema, et al. "Generative AI Misuse: A Taxonomy of Tactics and Insights from Real-World Data." *arXiv*, 19 June 2024, https://arxiv.org/abs/2406.13843.

10. Nissenbaum, Asaf, and Limor Shifman. "Internet Memes as Contested Cultural Capital: The Case of 4chan's /b/ Board." *New Media & Society*, vol. 19, no. 4, 2017, pp. 483–501, doi:10.1177/1461444815609313.

11. Papasavva, Antonis, et al. "Raiders of the Lost Kek: 3.5 Years of Augmented 4chan Posts from the Politically Incorrect Board." *arXiv [Cs.CY]*, 2020, http://arxiv.org/abs/2001.07487.

12. Sadasivam, Aadhavan, et al. "MemeBot: Towards Automatic Image Meme Generation." *arXiv [Cs.CL]*, 2020, http://arxiv.org/abs/2004.14571.

13. Siegel, Jacob. "Dylann Roof, 4chan, and the New Online Racism." *The Daily Beast*, 29 June 2015, https://www.thedailybeast.com/dylann-roof-4chan-and-the-new-online-racism\.

14. *Stylegan: StyleGAN - Official TensorFlow Implementation*. Accessed 20 Apr. 2025.

15. Zannettou, Savvas, et al. "On the Origins of Memes by Means of Fringe Web Communities." *arXiv [Cs.SI]*, 2018, http://arxiv.org/abs/1805.12512.

16. 4chan.org, https://www.4chan.org/. Accessed 20 Apr. 2025.

17. Adl.org, https://www.adl.org/resources/hate-symbol/pepe-frog. Accessed 20 Apr. 2025.
16